\documentclass[12pt]{article}

\newcommand{\jet}{{\rm jet}}

\newcommand{\pup}{p^{\uparrow}}

\newcommand{\GFt}{\widetilde{G}_F}

\newcommand{\rs}{\sqrt{S}}

\newcommand{\bea}{\begin{eqnarray}}
\newcommand{\eea}{\end{eqnarray}}
\newcommand{\nn}{\nonumber}

\newcommand{\bfc}{\begin{figure}\begin{center}}
\newcommand{\efc}{\end{center}\end{figure}}

\newcommand{\fig}[2]{\scalebox{#1}{\includegraphics{#2}}}

\newcommand{\GeV}{~{\rm GeV}}

\usepackage{graphicx}

\begin{document}

\vspace*{0.5cm}

\begin{center}

{\Large \bf Single transverse-spin asymmetry 
for direct-photon and single-jet productions at RHIC}

\vspace{1cm}

Koichi Kanazawa$^1$ and Yuji Koike$^2$

\vspace{1cm}

{\it $^1$ Graduate School of Science and Technology, Niigata University,
Ikarashi, \\ Niigata 950-2181, Japan}

\vspace{0.5cm}

{\it $^2$ Department of Physics, Niigata University,
Ikarashi, Niigata 950-2181, Japan}

\vspace{2.5cm}

{\large \bf Abstract} \end{center}

We study the
single transverse-spin asymmetry for the inclusive 
direct-photon 
and single-jet productions in the proton-proton collision based on
the twist-3 mechanism in the collinear factorization.
Taking into account all the effects from the twist-3 quark-gluon correlation functions
inside a transversely polarized proton,
we present a prediction for the asymmetries at the typical RHIC kinematics.  
In both processes
we find sizable asymmetries in the forward region of the polarized proton while they are almost zero in
the backward region.  
This implies that if one finds a nonzero asymmetries in the backward region
in these processes, it should be ascribed wholely to the three-gluon correlations.  
We also find the soft-gluon pole contribution is dominant
and the soft-fermion pole contribution
is negligible in the whole Feynman-$x$ region for these asymmetries.

\newpage

Study of large single transverse-spin asymmetry (SSA) in
inclusive reactions has
provided us with a range of new insights into the quark-gluon structure of
hadrons and has 
significantly developed the theoretical framework for the application of
perturbative QCD to hard
processes. (See \cite{D'AlesioMurgia2008,BaroneBradamanteMartin2010} for a review.)
When the transverse momentum of the final-state particle $P_T$ is large enough ($P_T\gg\Lambda_{\rm QCD}$), the SSA can
be described as a twist-3
observable
in the framework of the collinear factorization~\cite{EfremovTeryaev1982,QiuSterman1992,EguchiKoikeTanaka2007,
BeppuKoikeTanakaYoshida2010}.
From this perspective there have been many works which have explored the effects of twist-3 multiparton
correlation functions on
SSA~\cite{QiuSterman1992,QiuSterman1998,KanazawaKoike2000,KanazawaKoike2000E,KouvarisQiuVogelsangYuan2006,JiQiuVogelsangYuan2006DY,JiQiuVogelsangYuan2006SIDIS,KoikeVogelsangYuan2008,EguchiKoikeTanaka2006,KoikeTanaka2007M,KoikeTanaka2007,QiuVogelsangYuan2007,KoikeTomita2009,YuanZhou2009,KangYuanZhou2010,KanazawaKoike2010,KanazawaKoike2011,KanazawaKoike2011DY,KoikeTanakaYoshida2011,KoikeYoshida2011,KoikeYoshida2012,BeppuKoikeTanakaYoshida2012,KangMetzQiuZhou2011,KangQiuVogelsangYuan2011,KangProkudin2012}.
Among variety of observed SSAs, those for the inclusive single-hadron ($\pi$, $K$, $\eta$ etc) production
in the pp collision
at RHIC
\cite{Star2004,Phenix2005,Star2008,Brahms2008,Phenix2010,Star2012E,Star2012J}
are particularly suitable for the analysis based on the twist-3 mechanism,
since most data are in the range of $P_T\geq 1$ GeV and, in particular,
the next-to-leading order perturbative QCD calculation reproduces the twist-2 unpolarized cross 
section perfectly well~\cite{JagerSchaferStratmannVogelsang2003}.
In fact in the application of this mechanism to the 
RHIC $A_N$ data, 
it has been demonstrated that
the quark-gluon correlation function in the transversely polarized
proton reproduces the characteristic features of the observed 
asymmetry~\cite{KouvarisQiuVogelsangYuan2006,KanazawaKoike2010,KanazawaKoike2011}.  
This description, however, is based on the assumption 
that the whole asymmetry comes from the quark-gluon correlation.
Other sources of SSAs, such as the 
twist-3 fragmentation function~\cite{YuanZhou2009,KangYuanZhou2010} and the three-gluon correlation 
function~\cite{BeppuKoikeTanakaYoshida2010,KoikeTanakaYoshida2011,KoikeYoshida2011,KoikeYoshida2012},
may possibly bring a significant contribution to the asymmetry.

In order to clarify the origin of the observed SSA,
it is important to separate each competing effect by measuring SSAs in other processes.   
For example, the contribution from the twist-3
fragmentation function can be eliminated by studying 
direct-photon\footnote{Here we consider the isolated-photon production
in which the fragmentation contribution is suppressed by an appropriate isolation cut. } and single-jet
productions\footnote{The Drell-Yan process is another example
in which no fragmentation function contributes. 
But the hard-pole component of the quark-gluon correlation functions also contributes
to the cross section, which makes it difficult
to determine the form of each function~\cite{JiQiuVogelsangYuan2006DY,KanazawaKoike2011DY}.}
\bea
\pup + p \to \left\{
\begin{array}{l}
\gamma \\
\jet
\end{array}
\right\} + X. \label{process}
\eea
In these processes, the quark-gluon and the three-gluon correlation
functions bring asymmetries through two types of the pole contributions, i.e., 
soft-gluon-pole (SGP) and the soft-fermion
pole (SFP). 
For the direct-photon process, we have recently derived the contribution of
from the SFP component of the quark-gluon correlation function to the single-spin dependent
cross section at
leading-order (LO) perturbative QCD~\cite{KanazawaKoike2011DY}.
Combined with the contribution from the SGP 
component~\cite{QiuSterman1992,JiQiuVogelsangYuan2006DY,KoikeTanaka2007M} and the
three-gluon correlation function~\cite{KoikeYoshida2012}, the complete LO twist-3 formula
is currently available.
In principle, for these processes, one can see only the combined effect of the
quark-gluon and the three-gluon correlations.  
For the direct-photon production process, however, it has been shown that the three-gluon
correlation function does not give rise to $A_N$ 
at $x_F>0$ due to the smallness of the corresponding partonic cross section~\cite{KoikeYoshida2012}. 
Therefore, if a nonzero  $A_N^\gamma$ is experimentally observed at $x_F>0$, it should directly be
ascribed to the quark-gluon correlation function in the polarized nucleon.  
The SSA for the single-jet production also play a similar role in investigating
the multiparton correlations, although there is no
knowledge on the impact of the three-gluon correlation contribution at this point. 
Confrontation with future data for the processes
(\ref{process}) is particularly useful to test models for the quark-gluon correlation function
in the transversely polarized nucleon~\cite{KangQiuVogelsangYuan2011}.

The purpose of this Letter is to present a prediction for $A_N$ for the
processes (\ref{process}), using our model for the quark-gluon correlation functions
obtained in \cite{KanazawaKoike2010,KanazawaKoike2011}.  
There we have performed the fitting of the RHIC $A_N$ data for the $\pi$ and $K$ productions
based on the complete twist-3 cross section formula for the quark-gluon 
correlation functions, and have extracted the SGP and SFP components of those functions.  
The result reproduced all features of the observed asymmetries including those which were rather unexpected, 
such as the large $A_N$ for $K^-$ driven by the nonvalence component of the correlation function
and the peculiar $P_T$-dependence of the asymmetry, which had not been described by other analyses.  
In addition, the RHIC-STAR data for the $\eta$-meson
agreed with the {\it prediction} by the model~\cite{KanazawaKoike2011}, 
in which the strange-quark-gluon correlation responsible for
$A_N^{K^\pm}$ and the strangeness component in the $\eta$-meson fragmentation function play an important role.  
With these nice features at hand, the prediction of $A_N$ for (\ref{process}) 
will be useful as a reference for future experiment.  
One should keep in mind, however, that our model for the quark-gluon 
correlation functions was determined by assuming
that they are the sole origin for the observed $A_N$ for the light-hadron productions at RHIC.  
Therefore, if there is a discrepancy between our prediction and a future experiment, it would be a
signal for the existence of sizable twist-3 fragmentation or three-gluon correlation functions.  
We also remind that our SGP function does not agree with
what is expected from a naive relation between the SGP function and the moment of the Sivers function
(with respect to the transverse momentum $k_\perp$ of the quark) 
obtained from the analysis of the SSA data
of the semi-inclusive deep inelastic scattering~\cite{KangQiuVogelsangYuan2011,KangProkudin2012}. 
Here we put aside this issue\footnote{Clarification of this issue requires
the knowledge on the precise $k_\perp$-dependence in the high-$k_\perp$ region 
as well as the renormalization of the $k_\perp^2$-moment
of the Sivers function.}
and take the view of
investigating the prediction of
our model which can reproduce all the aspects of the
RHIC data for the light-hadron production.

We first recall some basic features of the quark-gluon
correlation contribution to the asymmetries for the direct-photon and the
single-jet productions.
The corresponding single-spin-dependent cross sections for these processes have
a common structure as
\cite{JiQiuVogelsangYuan2006DY,KoikeTanaka2007M,KanazawaKoike2011DY}
\bea
\Delta\sigma^{\gamma, \jet}
&\propto& \sum_{a,b} \left( G^a_{F}(x,x)-x\frac{dG^a_{F}(x,x)}{dx} \right) \otimes f^b(x')
\otimes \hat{\sigma}^{\rm SGP}_{ab\to \gamma, {\rm jet}} \nn\\
&+& \sum_{a,b} \left( G^a_{F}(0,x)+\widetilde{G}^a_{F}(0,x) \right)
\otimes f^b(x') \otimes
\hat{\sigma}^{\rm SFP}_{ab\to {\gamma, {\rm jet}}} , \label{qgc}
\eea
where $G_F^a$ and $\GFt^a$ are the quark-gluon correlation
functions for a quark or antiquark flavor $a$, 
$f^b(x')$ is the usual unpolarized parton
distribution function for the parton species $b$ ($b$=quark, antiquark or gluon). 
The symbol $\otimes$ denotes the
convolution with respect to the partonic momentum fraction $x$ and $x'$.  
$\hat{\sigma}_{ab\to\gamma,\jet}^{\rm SGP,SFP}$ represent the corresponding partonic hard cross
section for each subprocess and pole. 
The SGP and SFP functions, $G_F^a(x,x)$ and $G_F^a (0,x) + \GFt^a (0,x)$, for the light-quark flavors
($a=u,d,s,\bar{u},\bar{d},\bar{s}$) have been determined by an analysis
of the RHIC $A_N$ data for the inclusive pion
and kaon productions~\cite{KanazawaKoike2010}.
For the unpolarized parton distribution $f^b(x')$, we have
used the GRV98 LO parton distribution~\cite{GlueckReyaVogt1998}. 
Throughout this Letter, we choose the scales in the parton distribution and
fragmentation functions as $\mu=P_T$ as in the previous studies~\cite{KanazawaKoike2010,KanazawaKoike2011}.

\begin{figure}
 \begin{center}
  \fig{1.0}{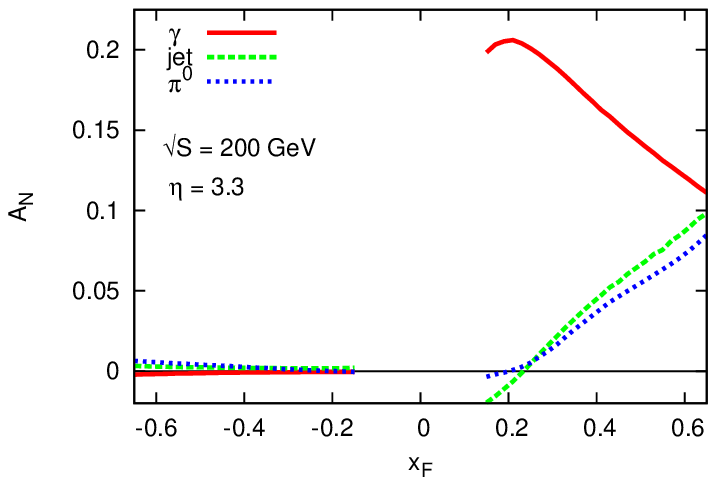}
  \fig{1.0}{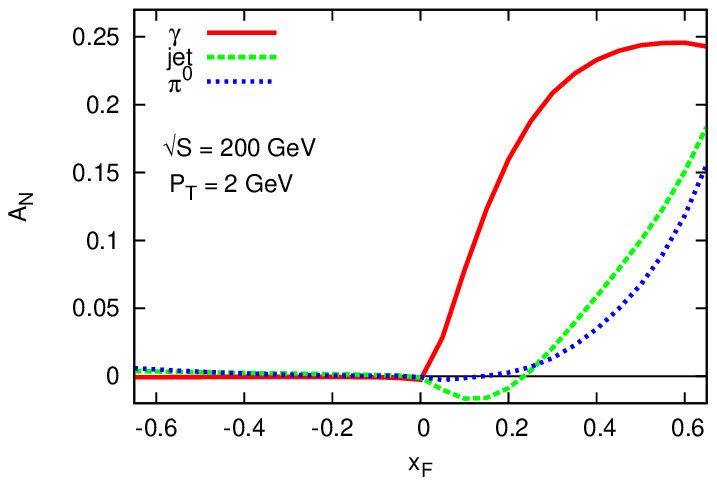}
 \end{center}
 \caption{Comparison of the $x_F$-dependence of $A_N$ for the direct-photon, jet  and $\pi^0$
productions at the center-of-mass energy $\sqrt{S}=200$ GeV. 
The left
 panel is for the fixed pseudorapidity $\eta=3.3$, while the right one is for the
 fixed transverse momentum $P_T=2\GeV$.  In the left panel, the plots are restricted in the
 region $P_T \ge 1$.  \label{200}}
\end{figure}

\begin{figure}
 \begin{center}
  \fig{1.0}{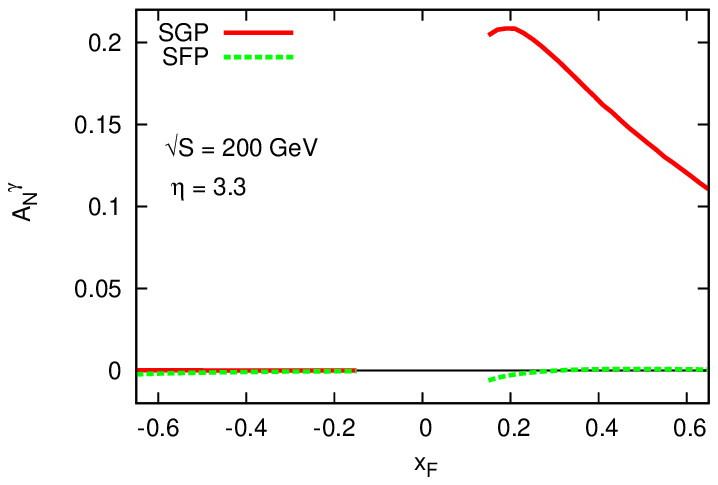}
  \fig{1.0}{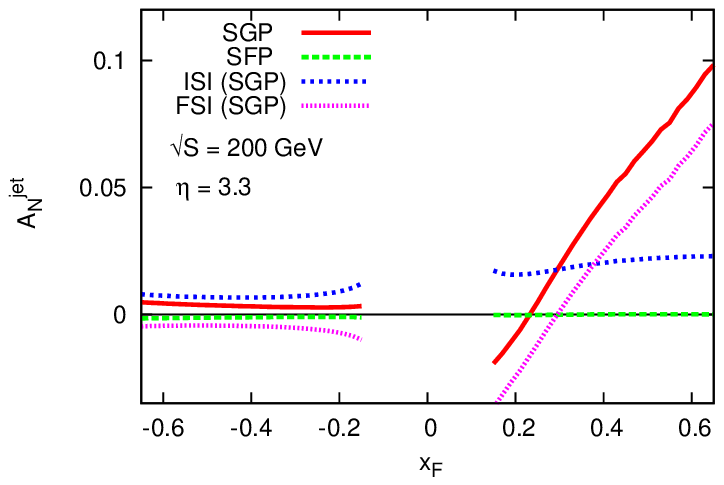}
 \end{center}
 \caption{Decomposition of $A_N^\gamma$ (left) and $A_N^\jet$ (right) in the left panel of Fig.~\ref{200} into the SGP and SFP
 contributions. Also
 shown in the right panel is the further decomposition of the SGP component into
 the initial-state and final-state interaction contributions.\label{bunri}}
\end{figure}

Figure \ref{200} shows $A_N$ for the direct-photon and the jet productions
as a function of $x_F$ at $\sqrt{S}=200\GeV$ at the pseudorapidity
$\eta=3.3$ and at $P_T=2\GeV$. In the figure we also plot
$A_N$ for the $\pi^0$ production for comparison.
First of all, $A_N^\gamma$ is significantly larger than $A_N^{\pi^0,{\rm jet}}$.
This is because the color factor for the polarized cross section relative to the unpolarized cross section
is much larger for the direct-photon production process than for the others.  
One also sees that the behavior of $A_N^\gamma$ is completely different from
$A_N^{\pi^0,{\rm jet}}$. 
As shown in the left panel of Fig. 1, for the fixed $\eta$, $A_N^{\gamma}$ has a peak at small $x_F$
and decreases as $x_F$ increases, while $A_N^{\pi^0,{\rm jet}}$ increases in the forward direction.  
These features at $x_F>0$ were also observed in the previous analysis of \cite{KangQiuVogelsangYuan2011}.
The similarity between $A_N^\jet$ and $A_N^{\pi^0}$ in their magnitude and behavior can be easily understood
because they have the common partonic subprocesses.

In order to see the origin of the different behavior between $A_N^\gamma$ and $A_N^{\rm jet}$, 
we first show the decomposition of $A_N^\gamma$ and $A_N^{\rm jet}$ into the SGP and SFP contributions in Fig.~\ref{bunri}.
From this decomposition it is clear that for these processes the dominant contribution is from
SGP and the effect of SFP is negligible in the whole $x_F$-region.  
(For the $K^-$ and $\pi^-$
productions, the SFP contribution survive as an important source of SSA in \cite{KanazawaKoike2010}.)
Therefore the above stated characteristic difference between $A_N^\gamma$ and $A_N^{\rm jet}$ is 
due to the difference in the SGP contribution to those asymmetries.
Next we
recall that the SGP cross section for $A_N^{\rm jet}$ consists of the initial-state 
interaction (ISI) and the final-state
interaction (FSI) 
contributions, and the rising behavior of $A_N^{\rm jet}$ toward large $x_F$ is mostly due to 
the latter one (See the right panel of Fig. \ref{bunri}.):  The partonic hard cross section for the latter
accompanies the kinematic factor $\hat{s}/\hat{t}$ compared to the former ($\hat{s}$ and $\hat{t}$ are
the Mandelstam variables in the parton
level) 
which enhances the asymmetry in the forward direction combined with
the derivative form of the SGP function.  
Since $A_N^\gamma$ receives the contribution only from the initial-state interaction,
it is not enhanced as $A_N^{\rm jet}$ in the forward direction.  
Another reason for the difference resides in the open partonic channels.
At moderate $x_F$, the channel $qg\to g\ (qg \to \gamma)$ gives rise to a major
contribution to $\Delta\sigma^\jet$ ($\Delta\sigma^\gamma$). 
At large $x_F$, however, the asymmetry 
for the jet production is dominated by the channel $qg\to q$, for which no counterpart exist in
the direct-photon production.  
As a consequence, $A_N^\gamma$ has a peak at moderate $x_F$ and decreases as a
function of $x_F$ in the forward region.
For fixed $P_T$, such behavior is softened as seen from the right panel of
Fig.~\ref{200}, and one finds $A_N^\gamma$ rapidly becomes zero with
$x_F\to 0$.

We remind that
an experimental observation of this characteristic feature of $A_N^\gamma$
requires a selection of only the direct-photon events.  Otherwise
the cross section for the prompt-photon production receives a large fragmentation
contribution.  The polarized cross section for the fragmentation contribution 
consists of the Sivers and Collins type contributions.  A recent model study shows that
the Collins type contribution is negligible compared with the Sivers type\,\cite{Gamberg:2012iq}.
Since the Sivers type contribution has the same partonic cross section as for the pion production, 
the asymmetry for the prompt-photon production 
which receives a large contamination from the fragmentation photon
would become closer to the one
for the pion and jet productions\,\cite{KouvarisQiuVogelsangYuan2006}.

Here we comment on the smallness of the SFP contribution in $A_N^{\gamma,{\rm jet}}$.  
As shown in~\cite{KanazawaKoike2011}, in the case of
the light-hadron productions, a sizable SFP
contribution to $A_N$
appears at moderate $x_F$ mostly through
the gluon fragmentation channels
owing to the large component of the gluon-to-pion and gluon-to-kaon fragmentation functions in the DSS
parametrization~\cite{FlorianSassotStratmann2007PIK} combined with the large 
partonic SFP cross section. 
For the direct-photon and the jet production processes, however, no such ``enhanced" 
SFP contribution exists because of the absence of the large gluon fragmentation
function.  Therefore $A_N^{\gamma,\jet}$ are a useful probe for the SGP component of the quark-gluon
correlation functions.  
As we saw in Fig. 1, for the direct-photon and the jet productions
the contribution from the quark-gluon correlation functions is almost zero
at $x_F<0$.  This means that if a future experiment finds nonzero
asymmetry at $x_F<0$, it should directly be ascribed to the three-gluon correlation functions in
the polarized nucleon.  
For the direct-photon process, it's been shown that the three-gluon correlation function
does not give rise to nonzero $A_N$ at $x_F>0$~\cite{KoikeYoshida2012}.  Thus
the quark-gluon correlation function studied in this Letter is the only source for $A_N$ in
this region. 
Accordingly measurement of $A_N^\gamma$ at both $x_F>0$ and $x_F<0$
gives an important information on both quark-gluon correlation and 
the three-gluon correlation in the polarized nucleon.

\begin{figure}
 \begin{center}
  \fig{1.0}{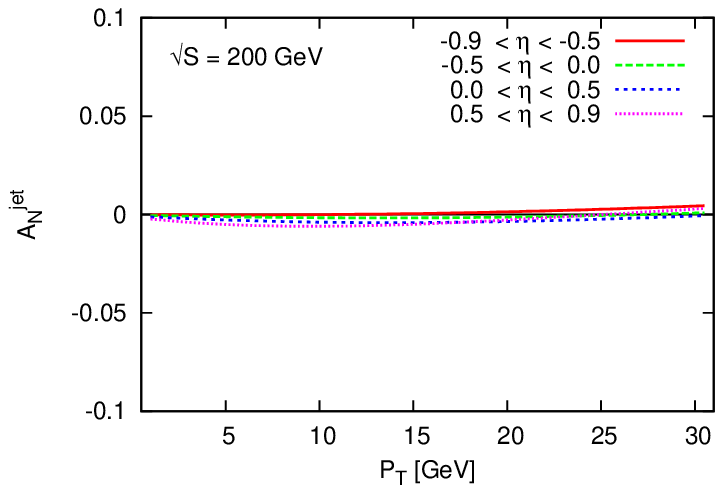}
  \fig{1.0}{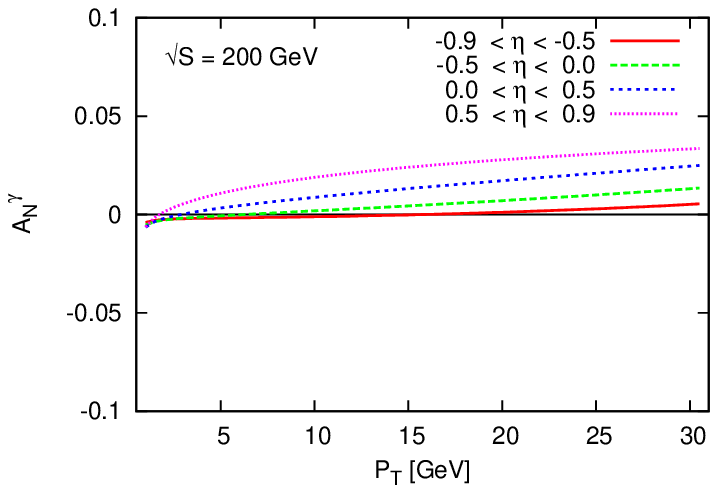}
 \end{center}
 \caption{$P_T$-dependence of $A_N^\jet$ and $A_N^\gamma$ at four different bins of the pseudo-rapidity $\eta$
 corresponding to the RHIC-STAR kinematics in \cite{Star2012J} \label{star}.}
\end{figure}

\begin{figure}
 \begin{center}
  \fig{1.0}{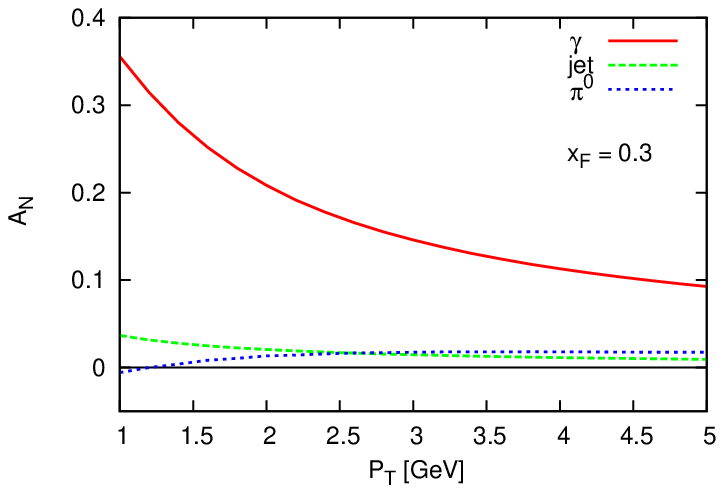}
  \fig{1.0}{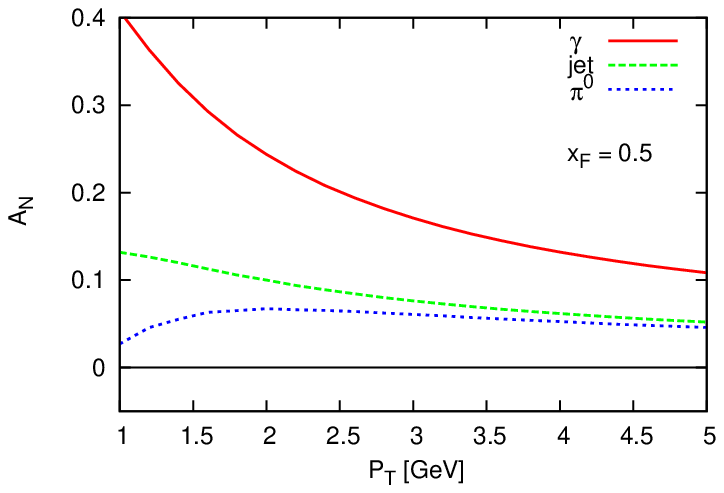}
 \end{center}
 \caption{Comparison of the $P_T$-dependence of $A_N$ for the direct-photon,
 jet, and $\pi^0$ productions at $\sqrt{S}=200\GeV$ for two different values of $x_F$. \label{pt}}
\end{figure}

Next, we explore the $P_T$-dependence of $A_N$, which gives an
important test for the twist-3 mechanism for SSA.  Shown in Fig. \ref{star} is the $P_T$-dependence of 
$A_N^{\rm jet}$ (left panel) and $A_N^\gamma$ (right panel)
at $\sqrt{S}=200$ GeV for several rapidity bins in the small $\eta$-region.  
One sees that $A_N^{\rm jet}$ is negligible at $0<P_T<30$ GeV, while
$A_N^\gamma$ is clearly finite, especially at $\eta>0$.  
For the jet production, the first data on the $P_T$-distribution
was recently reported by the STAR collaboration at mid-rapidity for $\sqrt{S}=200$ GeV~\cite{Star2012J}.
The data for $A_N^{\rm jet}$ is consistent with zero
with a large error bar for $0<P_T<30$ GeV, which agrees with 
the left panel of Fig. \ref{star}.  

For the direct-photon production, since our prediction shows finite $A_N^\gamma$
even at relatively small $\eta>0$,
measurement of $A_N^\gamma$ is expected to constrain the quark-gluon correlation function.
The $P_T$-dependence of $A_N^\gamma$ at forward-rapidity is more intriguing.
So far, data for the $P_T$-dependence in the forward region was
reported only for the inclusive $\pi^0$ production at $\sqrt{S}=200\GeV$ by the STAR collaboration~\cite{Star2008}, 
where the $P_T$-distribution has a peak at around a few GeV.
In our study~\cite{KanazawaKoike2011}, such peculiar
behavior of the $P_T$-dependence has been
reproduced owing to the large contributions from the gluon-fragmentation channel
to the polarized and the unpolarized cross sections with opposite signs,
both of which decrease quite fast as $P_T$ increases because of the fast evolution of the DSS
fragmentation function in the $P_T\sim$ a few GeV region.  
Owing to the absence of the fragmentation function
and the close relationship for the partonic cross sections between the unpolarized and 
the SGP cross sections, 
we expect that $A_N^{\gamma,{\rm jet}}$ approximately
follows the power behavior of the twist-3 asymmetry as
$A_N^{\gamma,{\rm jet}}\sim O\left({M_N/ P_T}\right)$
at {\it large} $x_F$.  
With this in mind, we show a comparison 
of the $P_T$-dependence of $A_N$ for the direct-photon, jet
and $\pi^0$ productions at two different values of $x_F$ in Fig.~\ref{pt}. 
As expected, $A_N^\gamma$ and $A_N^{\rm jet}$
decrease monotonously with increasing $P_T$
unlike the case for $\pi^0$. 
We found, however, that
the actual $A_N$ in Fig. \ref{pt} does not decrease
as fast as $1/P_T$ at large $P_T$,
which is due to 
the different functional forms of the SGP function and the unpolarized parton 
density\,\footnote{In this connection, we recall that the analysis in \cite{KanazawaKoike2010} adopted a
simplified scale-dependence for the quark-gluon correlation functions.  Use of the
correct scale-dependence for those functions\,\cite{Braun} may 
change this additional weak $P_T$-dependence.} as well
as the $P_T$-dependent phase space integral in the convolution.
Comparison of these features with a future measurement of 
the $P_T$-dependence of these asymmetries
in the forward region will shed new light on its validity.


\begin{figure}
 \begin{center}
  \fig{1.0}{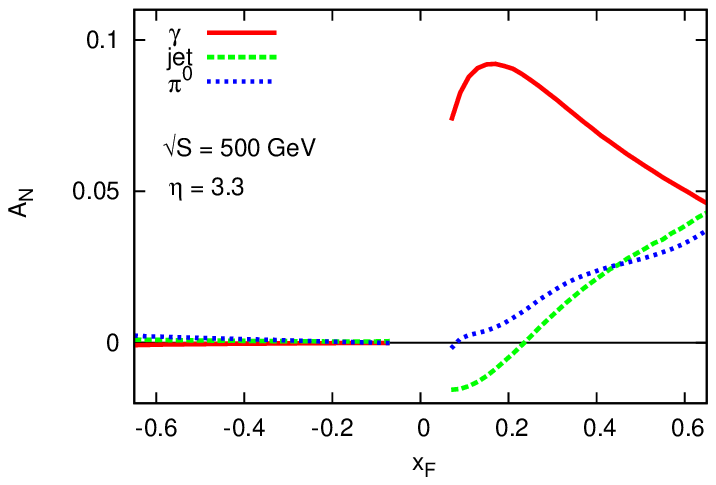}
  \fig{1.0}{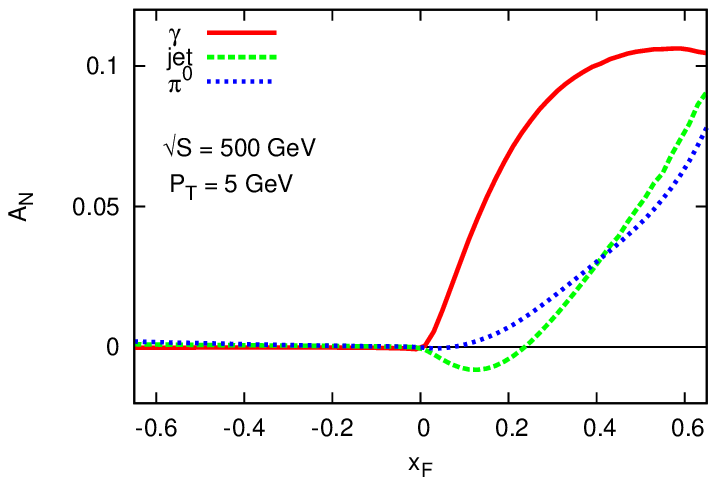}
 \end{center}
 \caption{$A_N$ at $\sqrt{S}=500\GeV$ for $\eta=3.3$ (left) and
 $P_T=5\GeV$ (right), respectively. \label{500}}
\end{figure}

Finally, to see the energy dependence of SSA, we show $A_N$ at higher energy, $\rs=500\GeV$, in
Fig.~\ref{500}.  From these plots, 
one finds the behavior of the asymmetry at $\rs=500\GeV$ is very similar to that at 
$\rs=200\GeV$ shown in Fig. 1, 
although the size of the asymmetry becomes  
approximately one-half of that for $\rs=200\GeV$.  
As in the case at $\sqrt{S}=200\GeV$, the dominant contribution
is from the SGP component 
in the whole $x_F$-region. 
Recently RHIC-A${}_N$DY collaboration reported a data for $A_N^{\rm jet}$ in the forward region
at $\sqrt{S}=500$ GeV\,\cite{Nogach2012}.  It shows a positive $A_N^{\rm jet}$ 
as shown in Fig. 5, but the magnitude is smaller. 
The difference in the magnitude may be ascribed to the fact 
that our quark-gluon correlation was determined so that it saturates the whole
asymmetry for the light-hadron production.  More data is needed to clarify the origin of the asymmetry.

In summary, we have presented a prediction for the SSA in the inclusive direct-photon and single-jet
productions for the typical RHIC kinematics, using the quark-gluon 
correlation function determined in our previous analysis of 
the light-hadron production.  
We have found $A_N^{\gamma}$ is significantly larger than
$A_N^{\pi^0}$ at moderate $x_F>0$, 
while the behavior of $A_N^{\rm jet}$ is similar to $A_N^{\pi^0}$.
In both processes, the SGP contribution dominates the asymmetry,
while the SFP contribution is negligible in the
whole $x_F$-region.  For the
direct-photon process, we have shown that
the asymmetries at $x_F>0$ and $x_F<0$ are, respectively, caused solely 
by the quark-gluon correlation function and the
three-gluon correlation function.  
These features of $A_N^\gamma$ and $A_N^{\rm jet}$ will
provide a unique opportunity for clarifying the mechanism of the observed asymmetries.

\section*{Acknowledgments}

The work of K.K. is supported by the Grand-in-Aid for
Scientific Research (No.24.6959) from the Japan Society
of Promotion of Science.
The work of Y.K. is supported in part by the Grant-in-Aid for
Scientific Research
(No.23540292) from the Japan Society of Promotion of Science.  


\end{document}